\documentclass[twocolumn,showpacs,superscriptaddress,amsmath,amssymb,prl]{revtex4-1}
\usepackage[normalem]{ulem}
\usepackage{epsfig}
\usepackage{amsfonts}
\usepackage{amsmath}
\usepackage{slashed}
\usepackage{graphicx}
\usepackage{color}
\usepackage{mathtools}
\usepackage{subfigure}
\usepackage{graphicx}
\usepackage{diagbox}

\usepackage{stmaryrd}
\usepackage{amssymb,psfrag}
\usepackage[normalem]{ulem}
\usepackage{caption}
\usepackage{multirow}

\newcommand{\beq}{\begin{eqnarray}}
\newcommand{\eeq}{\end{eqnarray}}

\newcommand{\nn}{\nonumber}

\begin{document}
\title{Linear relation between short range correlation and EMC effect of gluons in nuclei}

\author{Wei Wang}
\email{wei.wang@sjtu.edu.cn}
\affiliation{INPAC, Key Laboratory for Particle Astrophysics and Cosmology (MOE), Shanghai Key Laboratory for Particle Physics and Cosmology,
School of Physics and Astronomy, Shanghai Jiao Tong University, Shanghai 200240, China}

\author{Ji Xu}
\email{xuji\_phy@zzu.edu.cn}
\affiliation{School of Nuclear Science and Technology, Lanzhou University, Lanzhou 730000, China}
\affiliation{School of Physics and Microelectronics, Zhengzhou University, Zhengzhou, Henan 450001, China}

\author{Xing-Hua Yang}
\email{yangxinghua@sdut.edu.cn}
\affiliation{School of Physics and Optoelectronic Engineering, Shandong University of Technology, Zibo, Shandong 255000, China}

\author{Shuai Zhao}
\email{zhaos@tju.edu.cn}
\affiliation{Department of Physics, School of Science, Tianjin University, Tianjin 300350, China}

\date{\today}

\begin{abstract}
We explore  the EMC effect of gluons through heavy quark production in deep inelastic scattering (DIS) and  the nucleon-nucleon short range correlation (SRC)  from sub-threshold photoproduction of the $J/\psi$. Applying an effective field theory analysis, we scrutinize the gluon parton distribution functions  in nuclear environment and propose a linear relation between the slope of reduced cross section ratio, which reflects the magnitude of gluon EMC effect, and the  $J/\psi$ production cross section  at the photon energy $E_\gamma = 7.0\,\textrm{GeV}$ in the nucleus rest frame, a characteristic representative of SRC. This finding is supported by a numerical calculation using the available phenomenological results for gluon nuclear parton distribution functions (nPDFs).   Examining this linear relation by the forthcoming experimental data on electron-ion collider  allows to decipher nuclear effects of gluon distributions and enhance our comprehension of the substructure in bound nuclei.
\end{abstract}

\maketitle

\textit{Introduction.}---Atomic nuclei  that give rise to the vast majority of  mass in the visible universe are a system of strongly interacting nucleons (protons and neutrons). These nucleons are formed by constituent  quarks and gluons, with their characteristics   governed by principles of quantum chromodynamics (QCD). Despite the significant  advances in understanding the structure of nucleons, a sophisticated delineation  of the  substructure in bound nuclei is still lacking \cite{Hen:2016kwk}.

A clean and powerful platform to  decipher the substructure of nuclei is via  deep inelastic scattering process and many seminal  findings were made within the past decades. The renowned EMC effect \cite{EuropeanMuon:1983wih}, which reveals anomalous suppression of structure functions of atomic nuclei, is now widely regarded as an evidence that the momentum distributions of valence quarks in bound nucleons are modified compared to those of quarks in free nucleons at intermediate $0.35\leq x \leq 0.7$ \cite{EuropeanMuon:1988lbf,Gomez:1993ri,Seely:2009gt,Alde:1990im}. Experimental scrutiny of ratios of quasi-elastic (QE) scattering cross sections $a_2(A)=(\sigma_A(x,Q^2)/A)/(\sigma_d(x,Q^2)/2)$ at $1.5 \leq x \leq 2$  with $A$ the atomic number and $d$ the deuteron  points out that $a_2$ forms an $x$-independent plateau with negligible $Q^2$ dependence \cite{Frankfurt:1993sp,Fomin:2011ng,CLAS:2003eih,CLAS:2005ola,Hen:2014nza,CLAS:2018xvc,Li:2022fhh}. This is indicative of short range correlated pairs of interacting nucleons with a large relative momentum and a small center-of-mass (c.m.) momentum in comparison to the single-nucleon Fermi momentum, suggesting the presence of partons from the SRC pairs  with momenta surpassing the nucleonic average. Because of the relatively large motion of nucleon in SRCs, scattering at $x\geq1.5$ region can occur where scattering from a stationary proton is forbidden. The SRC scaling factor approximately equals the relative abundance of SRC pairs in a nucleus compared to deuteron. Recent research  advocate a tangible linear relationship between the EMC and SRC~\cite{Weinstein:2010rt,Hen:2012fm}, which leads to the conjectured description of the nuclei structure: for most of the time, nucleons bound in nuclei retain their free characteristics; while  for short time intervals nucleons undergo  fluctuations into SRC pairs, and thereby reshape the structures of nuclei substantially \cite{CLAS:2019vsb}.

However the mechanism that causes the modification of the valence quark momenta remains unknown, and yet there is no generally accepted explanation. An effective field theory (EFT) induced scale separation explanation has been proposed \cite{Chen:2016bde,Lynn:2019vwp}. Given the current circumstances, a natural question would be what role do gluons, the mediators of the strong force, play in the SRCs? Nucleons in pairs with short range correlation  can have denser local environment and larger momenta with high virtuality. These reconfigurations  are closely connected to the strong interactions in nuclei, and thereby the exploration of  distributions of mediators of strong interaction, namely gluons, in nuclei would be more incisive to validate this picture. In this work, we explore gluon EMC effect through heavy flavor production in DIS which is sensitive to the magnitude of the gluon EMC effect, and examine SRC  through sub-threshold photoproduction of $J/\psi$, a process  influenced  by  the SRC contributions. Based on an effective field theory (EFT) we propose a linear correspondence  between  the EMC effect and the SRC scaling factor measurements which is analogous to the quark domain. Utilizing parametrizations in EPPS21 and CT18ANLO \cite{Eskola:2021nhw,Hou:2019efy} as well as $a_2$ data in \cite{Hen:2012fm,CLAS:2019vsb}, we explore this linear relation in the gluon sector. Future experimental endeavors to probe this relation such as those at electron-ion collider (EIC)  can help to accomplish  the long standing quest for the nucleon substructure.

\textit{EMC and reduced cross section.}--Exploring gluon EMC effects necessitates an in-depth comprehension of the nuclear parton distribution functions (nPDFs) of gluons. To accurately delineate the gluon nPDF, one can undertake measurements of the production cross section of  heavy quark pair in electron-ion collision. The construction of EIC with a possibility to operate with wide variety of nuclei, will constrain the gluon density in nuclei via measurements of the charm reduced cross section \cite{Gluck:1987uk,Aschenauer:2017oxs}. For a given nucleus $A$, the charm reduced cross section is
\begin{eqnarray}\label{reduceCS}
  &&\sigma_{A,red}^{c \bar c}(x, Q^2) \equiv \left(\frac{d\sigma^{c \bar c}_A}{dx dQ^2}\right)\frac{xQ^4}{2\pi \alpha^2[1+(1-y)^2]} \nn\\
  &&\quad = \frac{2}{[1+(1-y)^2]}\left( xy^2 F_{1, A}^{c \bar c}(x,Q^2) + (1-y)F_{2, A}^{c \bar c}(x,Q^2) \right) \,,\nn\\
\end{eqnarray}
where $\sigma^{c \bar c}_A$ denotes the charm cross section which is customarily expressed as the reduced cross section $\sigma_{A,red}^{c \bar c}$; $F_{1/2, A}^{c \bar c}$ are charm structure functions depending on the gluon nPDFs.

\begin{figure}[htbp]
\includegraphics[width=1.00\columnwidth]{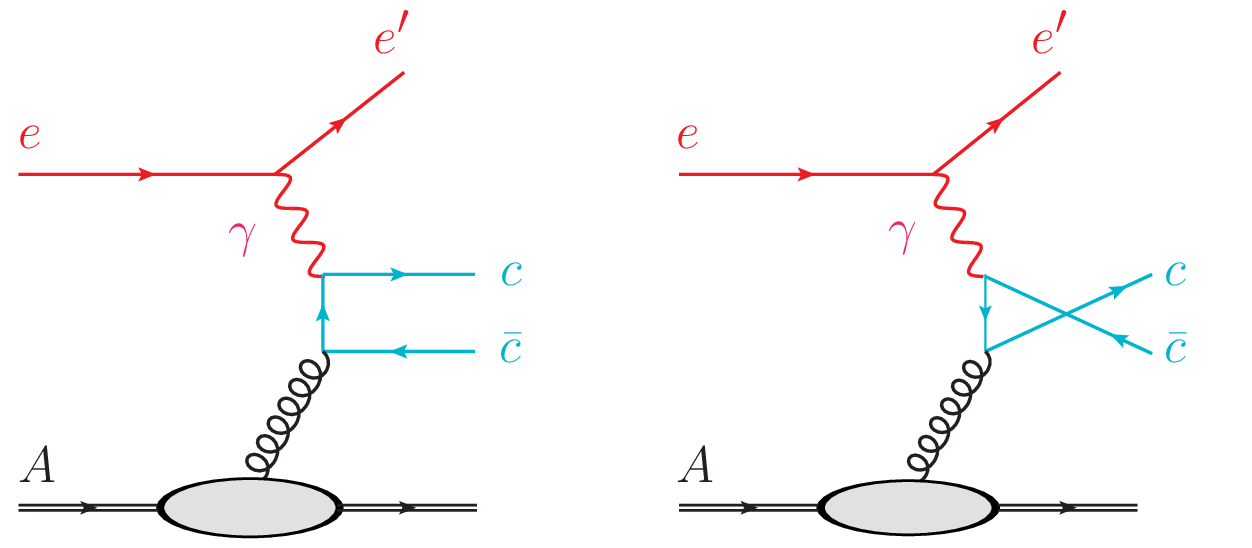}
\caption{Photoproduction of the heavy quark pair $c \bar c$  through photon-gluon fusion.}
\label{photon_gluon_fusion}
\end{figure}

Heavy quarks are predominantly generated through photon-gluon fusion due to their large masses, as shown in Figure \ref{photon_gluon_fusion}. The electron interacts with the target through a single virtual photon with momentum $\textbf{q}$ and energy $E_\gamma$, giving a four-momentum transfer $Q^2=|\textbf{q}|^2-E_\gamma^2$. Furthermore, the Bjorken scaling variable can be expressed as $x=Q^2/(2M_N E_\gamma)$, with $M_N$ denoting the nucleon mass. Additionally, the inelasticity $y$ represents the fraction of the lepton energy lost in the nucleon rest frame.

In the collinear factorization framework, one can express the charm structure function in terms of the perturbative  partonic cross section $f_{1/2}$ and the gluon nPDF $g_{A}$:
\begin{eqnarray}\label{final_SF12_integral_expression}
  F_{1,A}^{c \bar c}(x,Q^2) &=& \int_{\tau x}^1 \frac{dz}{z} g_A(z,\hat s) \, f_1(\frac{x}{z},Q^2) \,,\nn\\
  F_{2,A}^{c \bar c}(x,Q^2) &=& \int_{\tau x}^1 \frac{dz}{z} z \, g_A(z,\hat s) \, f_2(\frac{x}{z},Q^2) \,.
\end{eqnarray}
with $\tau \equiv 1+4M_c^2/Q^2$ and $\hat s=Q^2(\frac{z}{x}-1)$. The results of $f_{1/2}$ can be found in \cite{Laenen:1992zk}.

We utilize the results from global analysis of nPDFs presented by EPPS21 in which the gluon nPDFs $g_{A}(x, Q^2)$ can be expressed as the product of a factor $R_g^A$ and the gluon PDF in a free proton
\begin{eqnarray}\label{EPPS21}
  g_{A}(x, Q^2) = A R_g^A(x, Q^2) g(x, Q^2) \,.
\end{eqnarray}
The specific expression for $R_g^A$ can be found in \cite{Eskola:2021nhw}.

\begin{figure}[htbp]
\includegraphics[width=1.00\columnwidth]{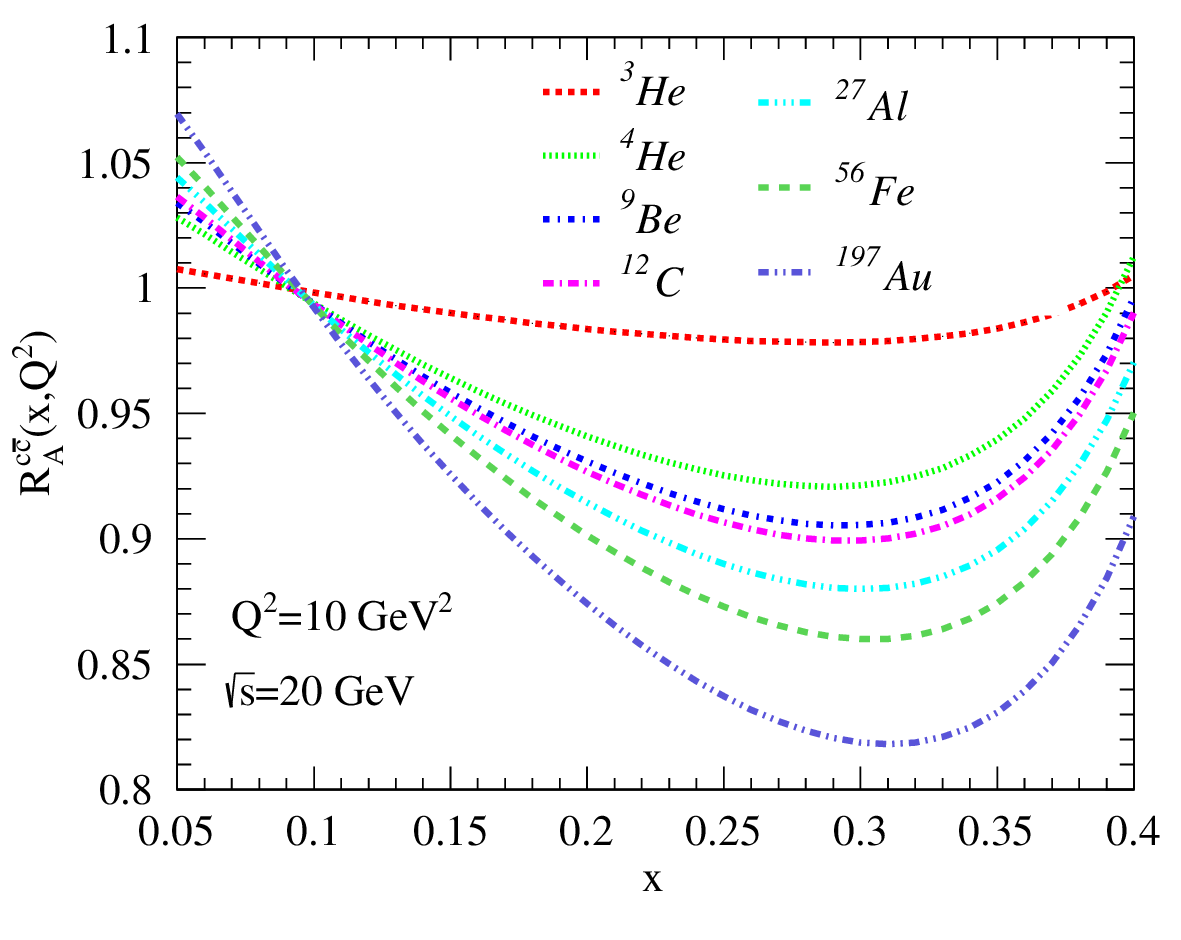}
\caption{$R_A^{c \bar{c}}\left(x, Q^2\right)$ defined in Eq.\,(\ref{nuclear_modi}) as a function of $x$,  and  $Q^2= 10\,\textrm{GeV}^2$, $\sqrt{s}=20\,\textrm{GeV}$. Different colours correspond to different nuclei, as indicated by the legends.}
\label{Rcc}
\end{figure}

We defined the factor $R_A^{c \bar{c}}$ to quantitatively assess the nuclear modification as the ratio below
\begin{eqnarray}\label{nuclear_modi}
  R_A^{c \bar{c}}\left(x, Q^2\right)=\frac{\sigma_{A,red}^{c \bar{c}}\left(x, Q^2\right)}{A \sigma_{N,red}^{c \bar{c}}\left(x, Q^2\right)},
\end{eqnarray}
where $\sigma_{N,red}^{c \bar{c}}\left(x, Q^2\right)$ denotes the reduced cross section in electron-proton collision. By utilizing Eqs.\,(\ref{reduceCS})$\sim$(\ref{nuclear_modi}), this ratio can be obtained. Figure \ref{Rcc} depicts $R_A^{c \bar{c}}$ as a function of $x$ with $A$ chosen to be $^{3}\textrm{He}$, $^{4}\textrm{He}$, $^{9}\textrm{Be}$, $^{12}\textrm{C}$, $^{27}\textrm{Al}$, $^{56}\textrm{Fe}$, and $^{197}\textrm{Au}$. The reason for selecting these nuclei is that they have been used for studying SRCs, thus the corresponding $a_2$ are available \cite{Hen:2012fm,CLAS:2019vsb}. Here $Q^2= 10\,\textrm{GeV}^2$ and $\sqrt{s}=20\,\textrm{GeV}$, which corresponds to the kinematics at the forthcoming EIC \cite{AbdulKhalek:2021gbh}.  One can find that the nuclear modification factor lies in the range  from 1.1 to 0.8 for $0.05\!\leq\! x \!\leq\! 0.3$. It is important to note that the actual momentum fraction carried by the gluon is larger than $x$ due to the convolutional form of the reduced cross section and the non-negligible charm quark mass. This can
be estimated as $x' \sim x \tau$. Consequently, the typical momentum fraction $x'$ spans approximately from 0.1 to 0.6 in this plot.


The magnitude of the gluon EMC effect for nucleus $A$ can be quantified by the slope of $R_A^{c \bar{c}}$, i.e., $|dR_A^{c \bar{c}}/dx|$. We fit this slope of the reduced cross section in the range of  $0.1 \leq x \leq 0.2$.
To incorporate the potential corrections, a systematic uncertainty is conducted by varying the kinematic range to $0.1 \leq x \leq 0.25$. The derived results are collected in the second column of Table \ref{slopeandsub}.


\textit{SRC and sub-threshold $J/\psi$ production.}---In the nuclei, nucleons interact in short ranges which can distort parton distributions. SRCs facilitate electron-ion scattering events at $x > 1$, a regime where interactions with free protons are prohibited. The correlated nucleons in pair have high relative but low c.m. momentum, in comparison to the Fermi momentum $k_F \leq 300\,\textrm{MeV}$, and the SRC scaling factor approximately equals the relative abundance of SRC pairs in a nucleus compared to a deuteron.

\begin{figure}[htbp]
\includegraphics[width=0.85\columnwidth]{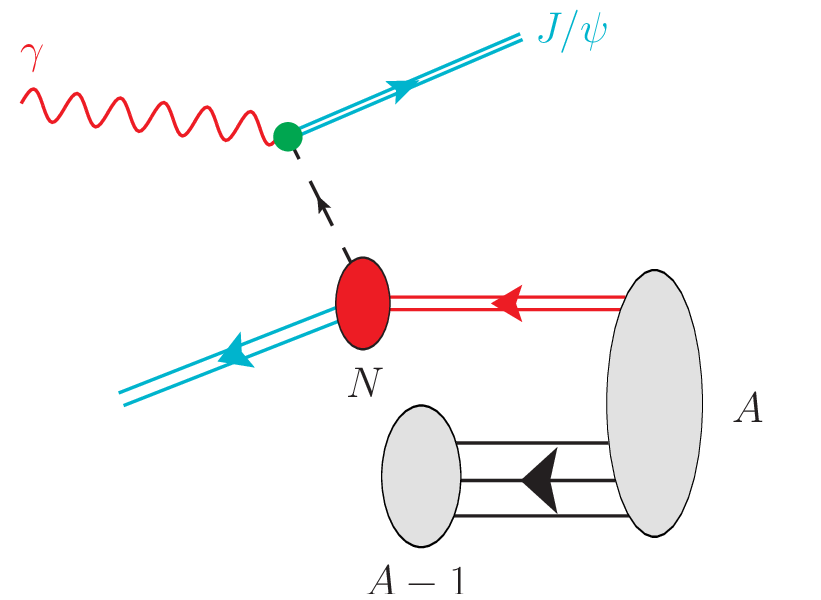}
\caption{$J/\psi$ production in $\gamma A$ collision. An incoming photon interacts with one nucleon $N$ from the nucleus $A$ and produces a $J/\psi$.}
\label{Schematic_diagram}
\end{figure}

The production of heavy quarkonia, specifically the $J/\psi$ meson in an electron-ion collision, which occurs effectively through $\gamma A$ interaction, enables the delineation of SRC phenomena. In such interactions, a minimum photon energy threshold of $E_\gamma \sim 8.2\,\textrm{GeV}$ in the nucleon rest frame is requisite for $J/\psi$ production when engaging a free nucleon. Contrastingly, in a nucleonic environment constrained within a nucleus, this threshold is reduced \cite{Brodsky:2000zc,Xu:2019wso}. In the deeply sub-threshold region ($E_\gamma \sim 7\,\textrm{GeV}$), the production of $J/\psi$ is particularly sensitive to the presence of SRC, positioning the $J/\psi$ production process as a prime diagnostic tool for SRC. This mechanism is depicted in Figure \ref{Schematic_diagram}.

The cross section per nucleon in $\gamma d$ collision is evaluated  as about $3.2\,\textrm{pb}$ \cite{Hatta:2019ocp}. Utilizing this finding in conjunction with the EFT analysis to follow, one can extrapolate the cross section per nucleon for different nuclei at the photon energy $E_\gamma \sim 7\,\textrm{GeV}$, with the quantitative outcomes presented subsequently.

\textit{Linear Gluon EMC-SRC relation from EFT.}---The fundamental origin of EMC and SRC remains enigmatic, with conjectures suggesting the existence of a linear relation in the quark sector. In the following we will perform an analysis based on chiral EFT and derive a linear relation between the magnitude of the gluon EMC effect and the sub-threshold $J/\psi$ production cross section. A similar analysis for the quark sector has been performed in~\cite{Chen:2016bde,Lynn:2019vwp}.

In the chiral EFT, the relevant scales below $Q$  are $\Lambda\sim 0.5$ GeV which is the range of validity of EFT,  and $P\sim m_{\pi}$ which is the typical momentum inside the nucleus. There is a hierarchy between these scales: $Q\gg \Lambda \gg P$. The QCD operators can be matched onto the EFT operators at $\Lambda$.
The Mellin moments of gluon nPDF are determined by matrix elements of the local operators
 \begin{eqnarray}
 	\langle A; p| \mathcal{O}_g^{\mu_0\cdots \mu_n} |A;p\rangle=	\langle x_g^n \rangle_A(Q^2) p^{(\mu_0}\cdots p^{\mu_n)} \,,
 \end{eqnarray}
 where
 \begin{eqnarray}
   \langle x_g^n \rangle_A(Q^2) = \int_{-A}^A x^n g_A(x, Q^2)dx \,, \nn
 \end{eqnarray}
and the local gauge invariant gluonic operators are
 \begin{eqnarray}
 	\mathcal{O}_g^{\mu_0\cdots\mu_n } = G_a^{\alpha(\mu_0}i D^{\mu_1}\cdots i D^{\mu_{n-1}} G_{a,\alpha}^{~~\mu_n)}.
 \end{eqnarray}
 They can be matched to hadronic operators in chiral EFT through
 \begin{eqnarray}\label{operatormatching}
 	\mathcal{O}_g^{\mu_0\cdots\mu_n }  &\to&  : \langle x_g^n \rangle_{N} M^n v^{(\mu_0}\cdots v^{\mu_n)}  N^{\dag} N \left( 1+\delta_n N^\dag N \right) \nn\\
 &&+\langle x_g^n \rangle_{\pi} \pi^{\alpha} i \partial^{(\mu_0}\cdots i \partial^{\mu_n)}\pi^{\alpha}+\cdots : \,.
 \end{eqnarray}
Here $(\cdots)$ indicates the indices within are symmetrized and traceless, and $:\,:$ denotes the normal ordering of the operator. $N, \pi$ are the nucleon and pion fields, $v$ is the nucleon four velocity.  $\langle x_g^n \rangle_{N(\pi)}$ is the $n$-th moment of gluon PDF in a free nucleon (pion). The term associated with $\delta_n$ contributes to the main EMC effect.

With the operator relation in Eq.\,\eqref{operatormatching}, the nuclear matrix element can be expressed as
\begin{eqnarray}\label{sepofxgnA}
  \langle x_g^n \rangle_{A}=A \langle x_g^n \rangle_{N}+ \langle x_g^n \rangle_{N}\delta_n(\Lambda,Q^2)  \langle A| \!:\!(N^{\dag}N)^2\!:\! |A\rangle \,,\nn\\
\end{eqnarray}
one can further derive a relation between the gluon PDFs by inverse Mellin transformation, which reads
\begin{eqnarray}\label{sepofgA}
  g_A(x, Q^2)/A \simeq g(x, Q^2) + g_2(A, \Lambda) \, \tilde{f}_g(x, Q^2, \Lambda) \,,
\end{eqnarray}
where the modification is depicted by the $A$-independent function $\tilde{f}_g$, and the magnitude of modification $g_2$, which represents the physics at scale $\Lambda$, is independent of $x$ and $Q^2$.
\begin{align}
  g_2(A,\Lambda)=\frac{1}{2A}  \langle A|(:N^{\dag}N:)^2 |A\rangle \,.
\end{align}
For the charm structure factors, according to Eq.\,(\ref{final_SF12_integral_expression}), one has
\begin{eqnarray}
  && \frac{F_{2,A}^{c \bar c}(x,Q^2)}{A} = \frac{1}{A}\int_{ax}^1 \frac{dz}{z} z \, g_A(z,\hat{s})f_2(\frac{x}{z},Q^2)\nonumber\\
  && \quad = \int_{ax}^1 \frac{dz}{z}  z \, g(z,\hat{s})  f_2(\frac{x}{z},Q^2) \nn\\
  &&\qquad +\int_{ax}^1 \frac{dz}{z} z g_2(A,\Lambda)\tilde{f}_g (z,\hat{s},\Lambda) f_2(\frac{x}{z},Q^2)\nonumber\\
  && \quad = F_{2, N}^{c \bar c}(x,Q^2) +g_2(A,\Lambda) \tilde{F}_2(x, Q,\Lambda) \,,
\end{eqnarray}
The $F_{2,N}^{c \bar c}$ refers to the charm structure function in a free nucleon. And
\begin{eqnarray}
  \tilde{F}_2(x, Q,\Lambda) \equiv \int_{ax}^1 \frac{dz}{z} z \tilde{f}_g (z,\hat{s},\Lambda) f_2(\frac{x}{z},Q^2)
\end{eqnarray}
is a function independent of $A$. Similarly, for $F_{1,A}^{c \bar c}$:
\begin{eqnarray}
  &&\frac{F_{1,A}^{c \bar c}(x,Q^2)}{A} = F_{1,N}^{c \bar c}(x,Q^2) +g_2(A,\Lambda) \tilde{F}_1(x, Q,\Lambda) \,, \nn\\
\end{eqnarray}
with
\begin{eqnarray}
  \tilde{F}_1(x, Q,\Lambda) \equiv \int_{ax}^1 \frac{dz}{z}  \tilde{f}_{g} (z,\hat{s},\Lambda) f_1(\frac{x}{z},Q^2) \,.
\end{eqnarray}
Therefore, one can also derive the relation for the reduced cross sections,
\begin{eqnarray}\label{eq:ffs}
  &&\frac{1}{A}\sigma_{A,red}^{c \bar c}(x, Q^2) = \frac{2}{[1+(1-y)^2]}\Big( xy^2 \frac{1}{A}F_{1, A}^{c \bar c}(x,Q^2) \nn\\
  && \qquad+ (1-y)\frac{1}{A}F_{2, A}^{c \bar c}(x,Q^2) \Big) \,,\nn\\
  &&\quad = \sigma_{red,N}^{c\bar c}
         (x,Q^2) +g_2(A,\Lambda) \tilde{\sigma}
 (x,Q^2,\Lambda) \,,
\end{eqnarray}
where
\begin{eqnarray}
  && \sigma_{red,N}^{c\bar c}(x,Q^2) = \nn\\
  &&\quad \frac{2}{[1+(1-y)^2]}\left( xy^2 F_{1, N}^{c \bar c}(x,Q^2) + (1-y)F_{2, N}^{c \bar c}(x,Q^2) \right) \,,\nn\\
  && \tilde{\sigma}(x,Q^2,\Lambda) = \nn\\
  &&\quad \frac{2}{[1+(1-y)^2]}\left( xy^2 \tilde{F}_{1}(x,Q^2) + (1-y)\tilde{F}_{2}(x,Q^2) \right) \,.\nn\\
\end{eqnarray}
It is worth noting that the function $\tilde{\sigma}(x,Q^2,\Lambda)$ is independent of $A$.

From the above results, the nuclear modification facotr $R_A^{c \bar{c}}$ can be expressed as
\begin{align}
  R_A^{c\bar c}(x,Q^2)=1+ g_2(A,\Lambda)\frac{\tilde{\sigma}(x,Q^2,\Lambda)}{\sigma_{red,N}^{c\bar c}(x,Q^2)} \,,
\end{align}
and for its derivative with $x$,
\begin{eqnarray}\label{LinearRelationEFT1}
\left|\frac{dR_A^{c \bar{c}}(x,Q^2)}{d x}\right| = C(x,Q^2) \, g_2(A,\Lambda) \,.
\end{eqnarray}
This shows a linear relation between the slope of nuclear modification $|dR_A^{c \bar{c}}/dx|$ and magnitude of modification $g_2$, with $C(x,Q^2)=\left| d \big(  \tilde{\sigma}(x,Q^2,\Lambda) / \sigma_{red, N}^{c \bar c}(x,Q^2)  \big)/dx \right|.$
Note that the separation of gluon nPDF indicates that the cross section of deeply sub-threshold $J/\psi$ production is dictated by the second term in Eq.\,(\ref{sepofgA}) where $\tilde{f}_g$ is universal among all nuclei. In this deeply sub-threshold region the SRCs dominate the cross section and $g_2$ indicates the magnitude of distortion induced by SRCs. Therefore, one can roughly link the ratio of cross sections of sub-threshold $J/\psi$ production for different nuclei to the ratio of $g_2$ through
\begin{eqnarray}\label{ratioofsubcross}
\frac{g_2(A, \Lambda)}{g_2(A', \Lambda)} \simeq   \left. \frac{\sigma_{A}^{sub}\!/A}{\sigma_{A'}^{sub}\!/A'} \right|_{E_{\gamma}\sim 7\,\textrm{GeV}} \,.
\end{eqnarray}
Choosing $A'$ as deuteron, Eq.\,(\ref{LinearRelationEFT1}) can be further developed into
\begin{eqnarray}\label{LinearRelationEFT2}
\left|\frac{dR_A^{c \bar{c}}(x,Q^2)}{d x}\right| = \left. \frac{C(x,Q^2) g_2(d,\Lambda)}{(\sigma_{d}^{sub}/2)} \, (\sigma_{A}^{sub}\!/A) \right|_{E_{\gamma}\sim 7\,\textrm{GeV}}  \,,\nn\\
\end{eqnarray}
This result draws a non-trivial prediction from chiral EFT on gluon nPDF. The linear relation can be examined with the experimental data of $\left|dR_A^{c \bar{c}}(x,Q^2)/d x\right| $ and ($\sigma_{A}^{sub}\!/A$) in future.

\textit{Test of the linear relation.}---To validate the linear relation, one needs the results of  $\left|dR_A^{c \bar{c}}(x,Q^2)/d x\right| $ and ($\sigma_{A}^{sub}\!/A$).
The experiment in Hall C at Jefferson Lab (JLab) has searched for the sub-threshold $J/\psi$ production from a carbon target at $E_\gamma \!\sim\! 5.7 \,\textrm{GeV}$ \cite{Bosted:2008mn}, but no event was observed. 
The EIC provides a unique opportunity to measure the cross section of sub-threshold $J/\psi$ production in the future.

Given the current scarcity of data, we employ two phenomenological strategies to approximate ($\sigma_{A}^{sub}\!/A$) and investigate the aforementioned linear relation. In the first strategy (I), we make use of the universality of the SRC scaling factor $a_2$, which is regarded as the relative probability for a nucleon in nucleus $A$ belongs to SRC pair compared to deuteron. It is postulated that $a_2$ is roughly equal to $g_2(A)/g_2(d)$ if the gluon nPDF is affected by the SRCs in about the same way as quarks.
Utilizing $a_2$ data from \cite{Hen:2012fm,CLAS:2019vsb} and the determined sub-threshold cross section of $(\sigma_d^{sub}/2)=3.2\,\textrm{pb}$ in $\gamma d$ collision \cite{Hatta:2019ocp}, we present the cross sections per nucleon for sub-threshold $J/\psi$ production at $E_\gamma \!\sim\! 7\,\textrm{GeV}$ for a range of nuclei in the third column of Table \ref{slopeandsub}, in which the uncertainties come from the uncertainties on measured $a_2$ in experiments. These values are denoted as $(\sigma_{A}^{sub}\!/A)^{\textrm{\uppercase\expandafter{\romannumeral1}}}$.

In the second strategy (II), we use Eq.\,(\ref{sepofgA}) to calculate the ratio of $g_2$ for different nuclei. From Eqs.\,\eqref{EPPS21} and \eqref{sepofgA},
\begin{eqnarray}\label{ratiofromEPPS21}
  g_2(A, \Lambda) \, \tilde{f}_g(x, Q^2, \Lambda) &\!\simeq\!& g_A(x, Q^2)/A - g(x, Q^2) \nn\\
  &\!=\!& \left( R_g^A(x, Q^2)\!-\!1 \right) g(x, Q^2) \,.
\end{eqnarray}
Therefore, one has
\begin{eqnarray}\label{ratiog2A}
  \frac{g_2(A, \Lambda)}{g_2(A', \Lambda)} = \frac{R_g^A(x, Q^2)-1}{R_g^{A'}(x, Q^2)-1} \,.
\end{eqnarray}
The left-hand side of Eq.\,(\ref{ratiog2A}) is independent of $x$ and $Q^2$, which necessitates a cancellation of these variables on the right-hand side in the EMC region, i.e., the $x$ and $Q^2$ dependent quantities combine to give an independent result.

\begin{widetext}
This characteristic has been corroborated with the fitted result of $R_g^A$ \cite{Eskola:2021nhw}.
Table \ref{ratiog2} presents the ratio $g_2(A)/g_2(^{12}$C) for different nuclei with $x\!=\!\{0.1, 0.2, 0.3, 0.4, 0.5, 0.6\}$ and $Q^2\!=\!\{10, 20, 30\}\,\textrm{GeV}^2$. One can observe that the values in each column change relatively small with variations in $x$ and $Q^2$.
\begin{table}[!htbp]
      \renewcommand{\arraystretch}{1.5}
      \caption{ The ratios of $g_2(A)/g_2(^{12}$C) for nuclei $^{3}$He, $^{4}$He, $^{9}$Be, $^{27}$Al, $^{56}$Fe and $^{197}$Au with respect to carbon $^{12}$C, at different Bjorken scale $x$ and $Q^2$. }\label{ratiog2}
	  \begin{tabular}{c|c|c|c|c|c|c}
		\hline\hline
		\diagbox[dir=NW]{$~~~~~~ (x,~Q^2)$}{$g_2(A)/g_2(^{12}$C)} & $^{3}$He & $^{4}$He & $^{9}$Be & $^{27}$Al & $^{56}$Fe & $^{197}$Au \\
		\hline
		$x=0.1, Q^2=10~{\rm GeV}^2$ & 0.21 & 0.78 & 0.94 & 1.20 & 1.41 & 1.87 \\
		\hline
		$x=0.1, Q^2=20~{\rm GeV}^2$ & 0.21 & 0.77 & 0.93 & 1.20 & 1.41 & 1.87 \\
		\hline	
		$x=0.1, Q^2=30~{\rm GeV}^2$ & 0.21 & 0.78 & 0.94 & 1.21 & 1.41 & 1.88 \\
		\hline
		$x=0.2, Q^2=10~{\rm GeV}^2$ & 0.21 & 0.75 & 0.95 & 1.27 & 1.49 & 2.02 \\
		\hline
		$x=0.2, Q^2=20~{\rm GeV}^2$ & 0.17 & 0.72 & 0.89 & 1.27 & 1.60 & 2.16 \\
		\hline
		$x=0.3, Q^2=10~{\rm GeV}^2$ & 0.23 & 0.83 & 0.95 & 1.15 & 1.30 & 1.64 \\
		\hline
		$x=0.3, Q^2=20~{\rm GeV}^2$ & 0.23 & 0.83 & 0.95 & 1.16 & 1.32 & 1.67 \\
		\hline	
		$x=0.3, Q^2=30~{\rm GeV}^2$ & 0.23 & 0.82 & 0.95 & 1.16 & 1.32 & 1.68 \\
		\hline
		$x=0.4, Q^2=10~{\rm GeV}^2$ & 0.23 & 0.82 & 0.95 & 1.16 & 1.33 & 1.69 \\
		\hline
		$x=0.4, Q^2=20~{\rm GeV}^2$ & 0.23 & 0.81 & 0.95 & 1.17 & 1.34 & 1.71 \\
		\hline	
		$x=0.4, Q^2=30~{\rm GeV}^2$ & 0.22 & 0.81 & 0.95 & 1.17 & 1.34 & 1.71 \\
		\hline
		$x=0.5, Q^2=10~{\rm GeV}^2$ & 0.22 & 0.81 & 0.95 & 1.17 & 1.34 & 1.70 \\
		\hline
		$x=0.5, Q^2=20~{\rm GeV}^2$ & 0.22 & 0.80 & 0.94 & 1.17 & 1.35 & 1.72 \\
		\hline	
		$x=0.5, Q^2=30~{\rm GeV}^2$ & 0.22 & 0.80 & 0.94 & 1.18 & 1.36 & 1.74 \\
		\hline
		$x=0.6, Q^2=10~{\rm GeV}^2$ & 0.22 & 0.80 & 0.94 & 1.18 & 1.36 & 1.74 \\
		\hline
		$x=0.6, Q^2=20~{\rm GeV}^2$ & 0.21 & 0.78 & 0.94 & 1.19 & 1.40 & 1.81 \\
		\hline	
		$x=0.6, Q^2=30~{\rm GeV}^2$ & 0.21 & 0.76 & 0.93 & 1.21 & 1.42 & 1.86 \\
		\hline\hline
     \end{tabular}
\end{table}	
\end{widetext}

Currently, there is no experimental data of sub-threshold production of $J/\psi$, nor is there any fitted result of deuteron in EPPS21. We use the calculated cross section of $^{12}$C from strategy (I) as input, $(\sigma_{\textrm{C}}^{sub}\!/12)^{\textrm{\uppercase\expandafter{\romannumeral1}}} = 14.37\pm0.54 \, \textrm{pb}$. Using this value as well as Eqs.\,(\ref{ratioofsubcross}) and (\ref{ratiog2A}), we have calculated the sub-threshold cross sections for other nuclei which are declared as $(\sigma_{A}^{sub}\!/A)^{\textrm{\uppercase\expandafter{\romannumeral2}}}$ and presented in the fourth column of Table \ref{slopeandsub}, in which the uncertainties come from $g_2(A)/g_2(^{12}$C) in each column in fitting Table \ref{ratiog2}.

\begin{table}[!htbp]
\renewcommand{\arraystretch}{1.5}
   \caption{The slopes of nuclear modifications $|dR_A^{c \bar{c}}/dx|$ as well as the cross sections (picobarn) per nucleon for sub-threshold $J/\psi$ production $(\sigma_{A}^{sub}\!/A)^{\textrm{\uppercase\expandafter{\romannumeral1}}}$ and $(\sigma_{A}^{sub}\!/A)^{\textrm{\uppercase\expandafter{\romannumeral2}}}$ obtained from two different phenomenological strategies. }\label{slopeandsub}
	\begin{tabular}{c|c c c}
		\hline\hline
		~~~Nucleus~~~ & ~~~$|dR_A^{c \bar{c}}/dx|$~~~ & ~~~$(\sigma_{A}^{sub}\!/A)^{\textrm{\uppercase\expandafter{\romannumeral1}}}$~~~ & ~~~$(\sigma_{A}^{sub}\!/A)^{\textrm{\uppercase\expandafter{\romannumeral2}}}$~~~ \\
		$^{3}$He   &  $0.135\pm0.010$   &  $6.82\pm0.13$   &  $3.31\pm0.06$  \\
		$^{4}$He   &  $0.495\pm0.035$   &  $11.52\pm0.32$  &  $11.35\pm0.14$  \\
        $^{9}$Be   &  $0.587\pm0.040$   &  $12.51\pm0.38$  &  $13.49\pm0.07$  \\
        $^{12}$C   &  $0.624\pm0.042$   &  $14.37\pm0.54$  &  $14.37\pm0.54$  \\
        $^{27}$Al  &  $0.738\pm0.049$   &  $15.46\pm0.58$  &  $17.23\pm0.23$ \\
        $^{56}$Fe  &  $0.857\pm0.056$   &  $15.36\pm0.70$  &  $20.09\pm0.42$ \\
        $^{197}$Au &  $1.106\pm0.073$   &  $16.51\pm0.70$  &  $26.13\pm0.82$  \\
		\hline\hline
	\end{tabular}
\end{table}	

\begin{figure}[htbp]
\centering
\includegraphics[width=1.00\columnwidth]{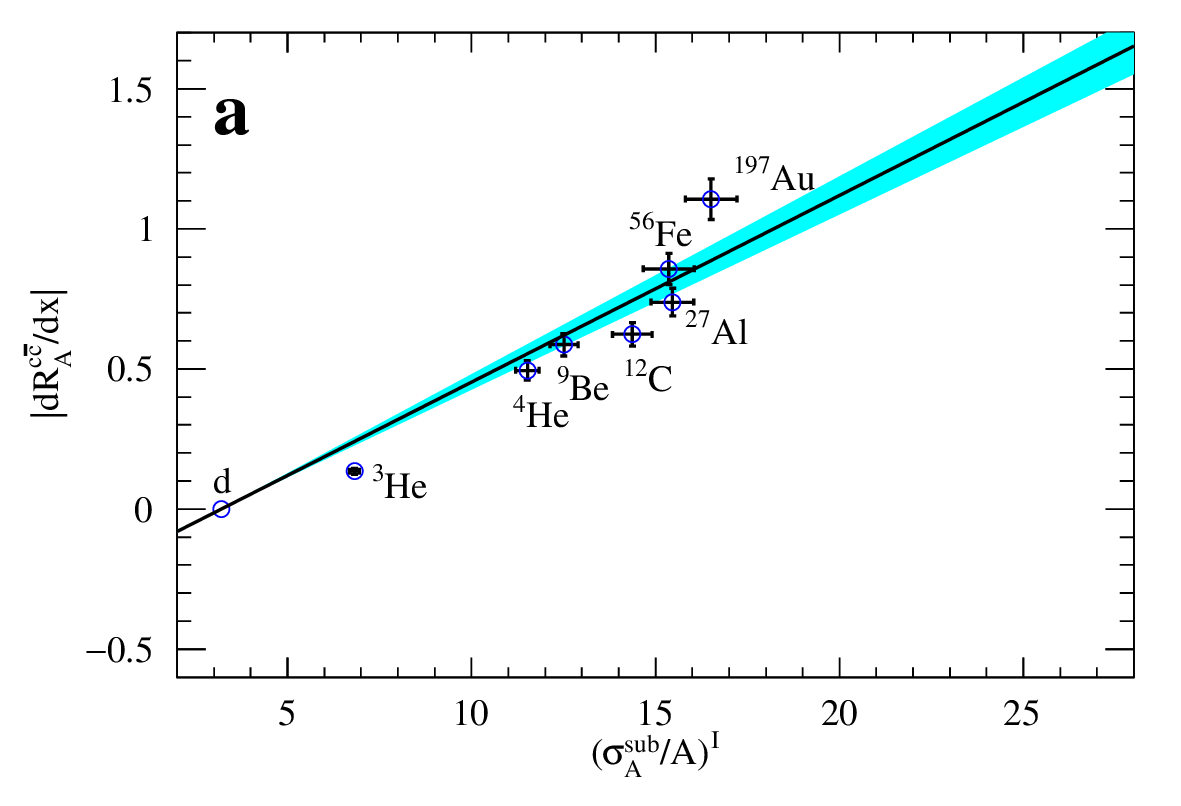}
\centering
\includegraphics[width=1.00\columnwidth]{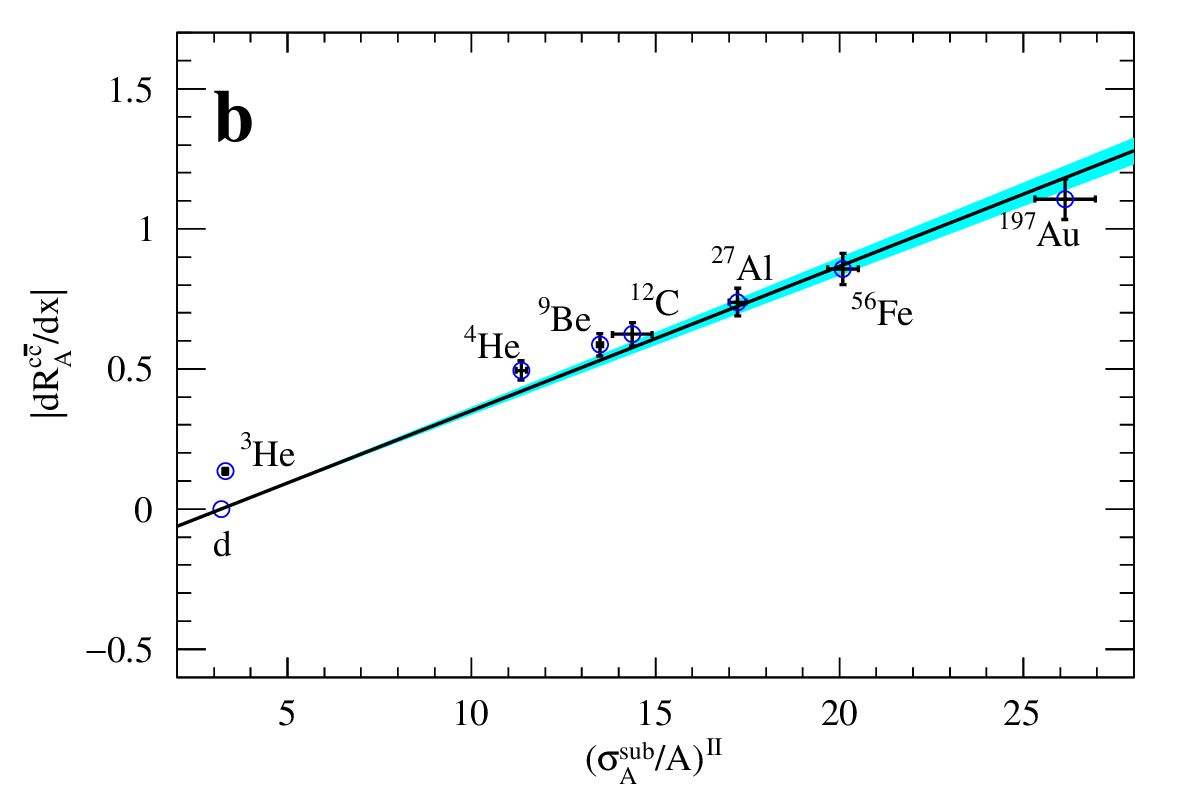}
\centering
\caption{The linear relation between the slope of nuclear modification $|dR_A^{c \bar{c}}/dx|$ and the sub-threshold cross section $(\sigma_{A}^{sub}\!/A)^{\textrm{\uppercase\expandafter{\romannumeral1}}}\textbf{(a)}$ or $(\sigma_{A}^{sub}\!/A)^{\textrm{\uppercase\expandafter{\romannumeral2}}}\textbf{(b)}$. The black line corresponds to a fit of numerical results, which is normalized to a reference point with $|dR_d^{c \bar{c}}/dx|=0$ and $(\sigma_{d}^{sub}/2)=3.2\,\textrm{pb}$.}
\label{LinearRelation}
\end{figure}

By analyzing the information delineated in Table~\ref{slopeandsub}, we plot the  gluon EMC slopes versus the cross sections for sub-threshold $J/\psi$ production in Figure \ref{LinearRelation}. The result reveals a linear relation between $|dR_A^{c \bar{c}}/dx|$ and $(\sigma_{A}^{sub}\!/A)$. The empirical relationship is encapsulated by
\begin{subequations}\label{LinearFit}
\begin{eqnarray}
  |d R_A^{c\bar c} / d x| &=& (0.067 \pm 0.004) \nn\\
  &&\times \left[ (\sigma_{A}^{sub}\!/A)^{\textrm{\uppercase\expandafter{\romannumeral1}}} - (\sigma^{sub}_d/2) \right] \,,\\
  |d R_A^{c\bar c} / d x| &=& (0.052 \pm 0.002) \nn\\
  &&\times \left[ (\sigma_{A}^{sub}\!/A)^{\textrm{\uppercase\expandafter{\romannumeral2}}} - (\sigma_d^{sub}/2) \right] \,.
\end{eqnarray}
\end{subequations}
This linear relation suggests that a common nuclear mechanism, likely the high virtuality nucleons in the nucleus, governs these two gluon-centric phenomena. We believe that although the linear relationship shown in Figure \ref{LinearRelation}(b) is better, the results presented in Figure \ref{LinearRelation}(a) are more persuasive, since the derivation of $(\sigma_{A}^{sub}\!/A)^{\textrm{\uppercase\expandafter{\romannumeral1}}}$ has nothing to do with the gluon nPDFs. In spite of the noticeable differences between the fitted slopes of two strategies, the linear association is apparent in both sets of fits in Eq.\,(\ref{LinearFit}). The chiral EFT provides us with a method for scale separation, which indicates that the modification part of gluon nPDF can be divided by two different quantities $g_2$ and $\tilde{f}_g$ with dependence on different energy scales, separately. If this separation holds, this linear relation would be a natural consequence. Nevertheless, the Chiral EFT can not tell us the properties on $\tilde{f}_g(x, Q^2, \Lambda)$, and the calculation of $g_2(A, \Lambda)$ would depend on assumptions and models. What we want to highlight is that the scale separation itself can provide us with some important information, like the existence of a linear relation, regardless of what the specific slope is.

The blue band in Figure \ref{LinearRelation} comes from fitting errors in Eq.\,(\ref{LinearFit}). It is crucial to note that the choice of model in this work will unavoidably introduce uncertainties and biases. Besides, our knowledge on gluon nPDF is very limited \cite{Duwentaster:2022kpv,AbdulKhalek:2022fyi,Detmold:2020snb}, and its error has not been taken into account. However, the introduction of nuclear modification $R_A^{c \bar{c}}$, as ratio of charm reduced cross sections for different nuclei, would reduce the corresponding uncertainties. In addition, the isoscalar correction dependencies of nPDFs are much milder in gluons than in quarks.

Notably, the extremities of the nuclear mass spectrum, specifically
$^{3}$He and $^{197}$Au, deviate from the predicted linear trend, potentially attributable to the very large proton-to-neutron ratio in $^{3}$He and the requirement for refined precision in heavy nuclei data \cite{Malace:2014uea}. The established linear relation in  Eq.\,(\ref{LinearFit}) facilitates the extrapolation of sub-threshold production for a wide range of nuclei such as $^{16}$O, $^{40}$Ca and $^{108}$Ag. Consequently, experimental validation of the linear relationship in Figure \ref{LinearRelation} is imperative. Additionally, ongoing investigations into the causal connection between the nucleons in SRCs and the EMC effect further accentuate the imperative for new experimental data \cite{Arrington:2012ax,Wang:2020uhj}.

\textit{Summary.}---
An examination of nuclear modification in gluon nPDFs is carried out through the study of heavy flavor production. In this work, we propose a linear correlation between the magnitude of the gluon EMC effect and the sub-threshold $J/\psi$ production cross section. This correlation provides a novel avenue  for verifying the nuclear modification of gluon distribution function and the universal influence of SRCs on parton distributions in nucleon.

The unambiguous signal presented by the initial discovery of EMC effect provided a portal to investigate QCD, a theory of quarks and gluons, in a cold nucleus. Despite the passage of four decades,  the EMC effect  remains an enigma to  both experimentalists and theorists. Nevertheless, the new experimental and theoretical results in recent years have reinvigorated the study of this longstanding conundrum. The sub-threshold production of $J/\psi$ is experimentally feasible at JLab and the future EIC \cite{Liu:2023htv}, our proposal in this work could be instrumental in finally resolving the origin of nuclear EMC effect.

\textit{Acknowledgements.}---This work is supported in part by National Natural Science Foundation of China under Grant No.  12125503, 12335003, 12105247, 12305106 and 12475098.


\begin{thebibliography}{}

\bibitem{Hen:2016kwk}
O.~Hen, G.~A.~Miller, E.~Piasetzky and L.~B.~Weinstein,
Rev. Mod. Phys. \textbf{89}, no.4, 045002 (2017)
doi:10.1103/RevModPhys.89.045002
[arXiv:1611.09748 [nucl-ex]].

\bibitem{EuropeanMuon:1983wih}
J.~J.~Aubert \textit{et al.} [European Muon],
Phys. Lett. B \textbf{123}, 275-278 (1983)
doi:10.1016/0370-2693(83)90437-9

\bibitem{EuropeanMuon:1988lbf}
J.~Ashman \textit{et al.} [European Muon],
Phys. Lett. B \textbf{202}, 603-610 (1988)
doi:10.1016/0370-2693(88)91872-2

\bibitem{Gomez:1993ri}
J.~Gomez, R.~G.~Arnold, P.~E.~Bosted, C.~C.~Chang, A.~T.~Katramatou, G.~G.~Petratos, A.~A.~Rahbar, S.~E.~Rock, A.~F.~Sill and Z.~M.~Szalata, \textit{et al.}
Phys. Rev. D \textbf{49}, 4348-4372 (1994)
doi:10.1103/PhysRevD.49.4348

\bibitem{Seely:2009gt}
J.~Seely, A.~Daniel, D.~Gaskell, J.~Arrington, N.~Fomin, P.~Solvignon, R.~Asaturyan, F.~Benmokhtar, W.~Boeglin and B.~Boillat, \textit{et al.}
Phys. Rev. Lett. \textbf{103}, 202301 (2009)
doi:10.1103/PhysRevLett.103.202301
[arXiv:0904.4448 [nucl-ex]].

\bibitem{Alde:1990im}
D.~M.~Alde, H.~W.~Baer, T.~A.~Carey, G.~T.~Garvey, A.~Klein, C.~Lee, M.~J.~Leitch, J.~W.~Lillberg, P.~L.~McGaughey and C.~S.~Mishra, \textit{et al.}
Phys. Rev. Lett. \textbf{64}, 2479-2482 (1990)
doi:10.1103/PhysRevLett.64.2479

\bibitem{Frankfurt:1993sp}
L.~L.~Frankfurt, M.~I.~Strikman, D.~B.~Day and M.~Sargsian,
Phys. Rev. C \textbf{48}, 2451-2461 (1993)
doi:10.1103/PhysRevC.48.2451

\bibitem{Fomin:2011ng}
N.~Fomin, J.~Arrington, R.~Asaturyan, F.~Benmokhtar, W.~Boeglin, P.~Bosted, A.~Bruell, M.~H.~S.~Bukhari, M.~E.~Christy and E.~Chudakov, \textit{et al.}
Phys. Rev. Lett. \textbf{108}, 092502 (2012)
doi:10.1103/PhysRevLett.108.092502
[arXiv:1107.3583 [nucl-ex]].

\bibitem{CLAS:2003eih}
K.~S.~Egiyan \textit{et al.} [CLAS],
Phys. Rev. C \textbf{68}, 014313 (2003)
doi:10.1103/PhysRevC.68.014313
[arXiv:nucl-ex/0301008 [nucl-ex]].

\bibitem{CLAS:2005ola}
K.~S.~Egiyan \textit{et al.} [CLAS],
Phys. Rev. Lett. \textbf{96}, 082501 (2006)
doi:10.1103/PhysRevLett.96.082501
[arXiv:nucl-ex/0508026 [nucl-ex]].

\bibitem{Hen:2014nza}
O.~Hen, M.~Sargsian, L.~B.~Weinstein, E.~Piasetzky, H.~Hakobyan, D.~W.~Higinbotham, M.~Braverman, W.~K.~Brooks, S.~Gilad and K.~P.~Adhikari, \textit{et al.}
Science \textbf{346}, 614-617 (2014)
doi:10.1126/science.1256785
[arXiv:1412.0138 [nucl-ex]].

\bibitem{CLAS:2018xvc}
M.~Duer \textit{et al.} [CLAS],
Phys. Rev. Lett. \textbf{122}, no.17, 172502 (2019)
doi:10.1103/PhysRevLett.122.172502
[arXiv:1810.05343 [nucl-ex]].

\bibitem{Li:2022fhh}
S.~Li, R.~Cruz-Torres, N.~Santiesteban, Z.~H.~Ye, D.~Abrams, S.~Alsalmi, D.~Androic, K.~Aniol, J.~Arrington and T.~Averett, \textit{et al.}
Nature \textbf{609}, no.7925, 41-45 (2022)
doi:10.1038/s41586-022-05007-2
[arXiv:2210.04189 [nucl-ex]].

\bibitem{Weinstein:2010rt}
L.~B.~Weinstein, E.~Piasetzky, D.~W.~Higinbotham, J.~Gomez, O.~Hen and R.~Shneor,
Phys. Rev. Lett. \textbf{106}, 052301 (2011)
doi:10.1103/PhysRevLett.106.052301
[arXiv:1009.5666 [hep-ph]].

\bibitem{Hen:2012fm}
O.~Hen, E.~Piasetzky and L.~B.~Weinstein,
Phys. Rev. C \textbf{85}, 047301 (2012)
doi:10.1103/PhysRevC.85.047301
[arXiv:1202.3452 [nucl-ex]].

\bibitem{CLAS:2019vsb}
B.~Schmookler \textit{et al.} [CLAS],
Nature \textbf{566}, no.7744, 354-358 (2019)
doi:10.1038/s41586-019-0925-9
[arXiv:2004.12065 [nucl-ex]].

\bibitem{Chen:2016bde}
J.~W.~Chen, W.~Detmold, J.~E.~Lynn and A.~Schwenk,
Phys. Rev. Lett. \textbf{119}, no.26, 262502 (2017)
doi:10.1103/PhysRevLett.119.262502
[arXiv:1607.03065 [hep-ph]].

\bibitem{Lynn:2019vwp}
J.~E.~Lynn, D.~Lonardoni, J.~Carlson, J.~W.~Chen, W.~Detmold, S.~Gandolfi and A.~Schwenk,
J. Phys. G \textbf{47}, no.4, 045109 (2020)
doi:10.1088/1361-6471/ab6af7
[arXiv:1903.12587 [nucl-th]].

\bibitem{Eskola:2021nhw}
K.~J.~Eskola, P.~Paakkinen, H.~Paukkunen and C.~A.~Salgado,
Eur. Phys. J. C \textbf{82}, no.5, 413 (2022)
doi:10.1140/epjc/s10052-022-10359-0
[arXiv:2112.12462 [hep-ph]].

\bibitem{Hou:2019efy}
T.~J.~Hou, J.~Gao, T.~J.~Hobbs, K.~Xie, S.~Dulat, M.~Guzzi, J.~Huston, P.~Nadolsky, J.~Pumplin and C.~Schmidt, \textit{et al.}
Phys. Rev. D \textbf{103}, no.1, 014013 (2021)
doi:10.1103/PhysRevD.103.014013
[arXiv:1912.10053 [hep-ph]].

\bibitem{Gluck:1987uk}
M.~Gluck, R.~M.~Godbole and E.~Reya,
Z.\ Phys.\ C {\bf 38}, 441 (1988)
Erratum: [Z.\ Phys.\ C {\bf 39}, 590 (1988)].
doi:10.1007/BF01584394

\bibitem{Aschenauer:2017oxs}
E.~C.~Aschenauer, S.~Fazio, M.~A.~C.~Lamont, H.~Paukkunen and P.~Zurita,
Phys. Rev. D \textbf{96}, no.11, 114005 (2017)
doi:10.1103/PhysRevD.96.114005
[arXiv:1708.05654 [nucl-ex]].

\bibitem{Laenen:1992zk}
E.~Laenen, S.~Riemersma, J.~Smith and W.~L.~van Neerven,
Nucl. Phys. B \textbf{392}, 162-228 (1993)
doi:10.1016/0550-3213(93)90201-Y

\bibitem{AbdulKhalek:2021gbh}
R.~Abdul Khalek, A.~Accardi, J.~Adam, D.~Adamiak, W.~Akers, M.~Albaladejo, A.~Al-bataineh, M.~G.~Alexeev, F.~Ameli and P.~Antonioli, \textit{et al.}
Nucl. Phys. A \textbf{1026}, 122447 (2022)
doi:10.1016/j.nuclphysa.2022.122447
[arXiv:2103.05419 [physics.ins-det]].

\bibitem{Brodsky:2000zc}
S.~J.~Brodsky, E.~Chudakov, P.~Hoyer and J.~M.~Laget,
Phys. Lett. B \textbf{498}, 23-28 (2001)
doi:10.1016/S0370-2693(00)01373-3
[arXiv:hep-ph/0010343 [hep-ph]].

\bibitem{Xu:2019wso}
J.~Xu and F.~Yuan,
Phys. Lett. B \textbf{801}, 135187 (2020)
doi:10.1016/j.physletb.2019.135187
[arXiv:1908.10413 [hep-ph]].

\bibitem{Hatta:2019ocp}
Y.~Hatta, M.~Strikman, J.~Xu and F.~Yuan,
Phys. Lett. B \textbf{803}, 135321 (2020)
doi:10.1016/j.physletb.2020.135321
[arXiv:1911.11706 [hep-ph]].


\bibitem{Bosted:2008mn}
P.~Bosted, J.~Dunne, C.~A.~Lee, P.~Junnarkar, M.~Strikman, J.~Arrington, R.~Asaturyan, F.~Benmokhtar, M.~E.~Christy and E.~Chudakov, \textit{et al.}
Phys. Rev. C \textbf{79}, 015209 (2009)
doi:10.1103/PhysRevC.79.015209
[arXiv:0809.2284 [nucl-ex]].

\bibitem{Duwentaster:2022kpv}
P.~Duwent\"aster, T.~Je\v{z}o, M.~Klasen, K.~Kova\v{r}\'\i{}k, A.~Kusina, K.~F.~Muzakka, F.~I.~Olness, R.~Ruiz, I.~Schienbein and J.~Y.~Yu,
Phys. Rev. D \textbf{105}, no.11, 114043 (2022)
doi:10.1103/PhysRevD.105.114043
[arXiv:2204.09982 [hep-ph]].

\bibitem{AbdulKhalek:2022fyi}
R.~Abdul Khalek, R.~Gauld, T.~Giani, E.~R.~Nocera, T.~R.~Rabemananjara and J.~Rojo,
Eur. Phys. J. C \textbf{82}, no.6, 507 (2022)
doi:10.1140/epjc/s10052-022-10417-7
[arXiv:2201.12363 [hep-ph]].

\bibitem{Detmold:2020snb}
W.~Detmold \textit{et al.} [NPLQCD],
Phys. Rev. Lett. \textbf{126}, no.20, 202001 (2021)
doi:10.1103/PhysRevLett.126.202001
[arXiv:2009.05522 [hep-lat]].

\bibitem{Malace:2014uea}
S.~Malace, D.~Gaskell, D.~W.~Higinbotham and I.~Cloet,
Int. J. Mod. Phys. E \textbf{23}, no.08, 1430013 (2014)
doi:10.1142/S0218301314300136
[arXiv:1405.1270 [nucl-ex]].

\bibitem{Arrington:2012ax}
J.~Arrington, A.~Daniel, D.~Day, N.~Fomin, D.~Gaskell and P.~Solvignon,
Phys. Rev. C \textbf{86}, 065204 (2012)
doi:10.1103/PhysRevC.86.065204
[arXiv:1206.6343 [nucl-ex]].

\bibitem{Wang:2020uhj}
X.~G.~Wang, A.~W.~Thomas and W.~Melnitchouk,
Phys. Rev. Lett. \textbf{125}, 262002 (2020)
doi:10.1103/PhysRevLett.125.262002
[arXiv:2004.03789 [hep-ph]].

\bibitem{Liu:2023htv}
T.~Liu, Z.~W.~Zhao, M.~Cai, D.~Byer and H.~Gao,
[arXiv:2310.05405 [nucl-ex]].










\end{thebibliography}
\end{document}